\documentclass[9pt, journal]{IEEEtran}
\newcommand{\subparagraph}{}
\usepackage{cite}
\usepackage{amsmath,amssymb,amsfonts}
\pdfoutput=1
\usepackage{graphicx}
\usepackage{caption}
\usepackage{comment}
\usepackage[ruled,vlined]{algorithm2e}
\include{pythonlisting}
\usepackage{float}
\usepackage{graphics}
\usepackage{titlesec}
\usepackage{subfig}
\usepackage{graphicx}
\usepackage{xcolor}
\usepackage{titlesec}
\usepackage{subfig}
\usepackage{graphicx}
\usepackage{mathtools}
\usepackage{cite}
\usepackage[font=footnotesize]{caption}
\usepackage[english]{babel}
\usepackage[utf8x]{inputenc}
\usepackage{amsmath}
\usepackage{xcolor}
\usepackage{flushend}
\usepackage[colorinlistoftodos]{todonotes}

\usepackage{enumitem}

\titlespacing{\section}{0pt}{2ex}{1ex}
\titlespacing{\subsection}{0pt}{0.5ex}{0.2ex}
\titlespacing{\subsubsection}{0pt}{0.5ex}{0ex}
\begin{document}
\title{ENOS: Energy-Aware Network Operator Search for Hybrid Digital and Compute-in-Memory DNN Accelerators}
\author{Shamma Nasrin, Ahish Shylendra, Yuti Kadakia, Nick Iliev, Wilfred Gomes, Theja Tulabandhula, and Amit Ranjan Trivedi}

\maketitle
\begin{abstract}
This work proposes a novel \textit{Energy-Aware Network Operator Search (ENOS)} approach to address the energy-accuracy trade-offs of a deep neural network (DNN) accelerator. In recent years, novel inference operators such as binary weight, multiplication-free, and deep shift have been proposed to improve the computational efficiency of a DNN. Augmenting the operators, their corresponding novel computing modes such as compute-in-memory and XOR networks have also been explored. However, simplification of DNN operators invariably comes at the cost of lower accuracy, especially on complex processing tasks. While the prior works explore an end-to-end DNN processing with the same operator and computing mode, our proposed ENOS framework allows an \textit{optimal layer-wise integration of inference operators and computing modes} to achieve the desired balance of energy and accuracy. The search in ENOS is formulated as a continuous optimization problem, solvable using typical gradient descent methods, thereby scalable to larger DNNs with minimal increase in training cost. We characterize ENOS under two settings. In the first setting, for digital accelerators, we discuss ENOS on multiply-accumulate (MAC) cores that can be reconfigured to different operators. ENOS training methods with single and bi-level optimization objectives are discussed and compared. We also discuss a sequential operator assignment strategy in ENOS that only learns the assignment for one layer in one training step, enabling greater flexibility in converging towards the optimal operator allocations. Furthermore, following Bayesian principles, a sampling-based variational mode of ENOS is also presented. ENOS is characterized on popular DNNs ShuffleNet and SqueezeNet on CIFAR10 and CIFAR100. Compared to the conventional uni-operator approaches, under the same energy budget, ENOS improves accuracy by 10--20\%. In the second setting, for a hybrid digital and compute-in-memory accelerator, we characterize ENOS to assign both layer-wise computing mode (high precision digital or low precision analog compute-in-memory) as well as operator while staying within the total compute-in-memory budget. Under varying configurations of hybrid accelerator, ENOS can leverage higher energy efficiency of compute-in-memory operations to reduce the operating energy of DNNs by 5$\times$ while suffering $<$1\% reduction in accuracy. Characterization results using ENOS show interesting insights, such as amenability of different filters to using low complexity operators, minimizing the energy of inference while maintaining high prediction accuracy.     

\end{abstract}
  
\section{Introduction}
The proliferation of deep learning in embedded computing space has aggravated computational efficiency challenges of deep neural networks (DNNs). Consequently, DNN architectures are going through a dramatic evolution to improve their prediction capacity within limited computing and storage resources. While so far, DNNs were hand-designed by domain experts, increasingly, more attention is being dedicated to automating their architecture search. This sub-field of machine learning, called Neural Architecture Search (NAS), has seen rapid growth. Among recent NAS approaches, NASnet\cite{DBLP} learns the best convolutional architecture on a smaller dataset using reinforcement learning and then transfers the learned architecture to a larger dataset. In Efficient Neural Architecture Search (ENAS) \cite{Efficient}, a controller is trained using policy gradient to find an optimal subgraph within a large computational graph. Unlike prior reinforcement learning-based approaches, which learn over a discrete search space and thereby require thousands of GPU hours to compute, in \cite{DART}, architecture search was solved using gradient descent by formulating the search space to be continuous. Using \cite{DART}, the architecture search could be performed within a few GPU hours on modern DNNs.

In parallel, considerable efforts have also focused on optimizing DNN's correlation operators. Scalar multiplication of input and weight vectors is the most prevalent correlation operator in DNNs. Since DNN's processing involves evaluating thousands of correlations among high-dimensional input/weight vectors, the operator dictates the overall computational efficiency of a DNN. Therefore, prior works have explored alternate, more computationally efficient correlation operators. \cite{MFNET,date} introduced multiplication-free operation where high-precision multiplication between input and weight is replaced by 1-bit multiplication and multi-bit addition. \cite{deepShift} introduced bit-wise shift operators where weights are approximated by a power of two, i.e., 2$^n$ [$n \in \mathbb{Z}^+$], replacing multiplications with shift operations. \cite{courbariaux2016binarized, binaryconect} quantize inputs and weights to one-bit and uses XNOR instead of multiplications. The new correlation operators further benefit from the custom-designed computing modes. For example, multiplication-free and binary weights operators are quite suited for compute-in-memory \cite{MFNET,nasrin2020supported,shukla2020mc}. Deep-shift operator in \cite{deepShift} can be implemented using digital shifters. Operators in \cite{courbariaux2016binarized, binaryconect} can be efficiently implemented using XOR logic gates. Combining the operators with an optimal computing mode im
proves the energy efficiency of DNNs considerably. 

\begin{figure}[!t]
     \centering
     \includegraphics[width=\linewidth]{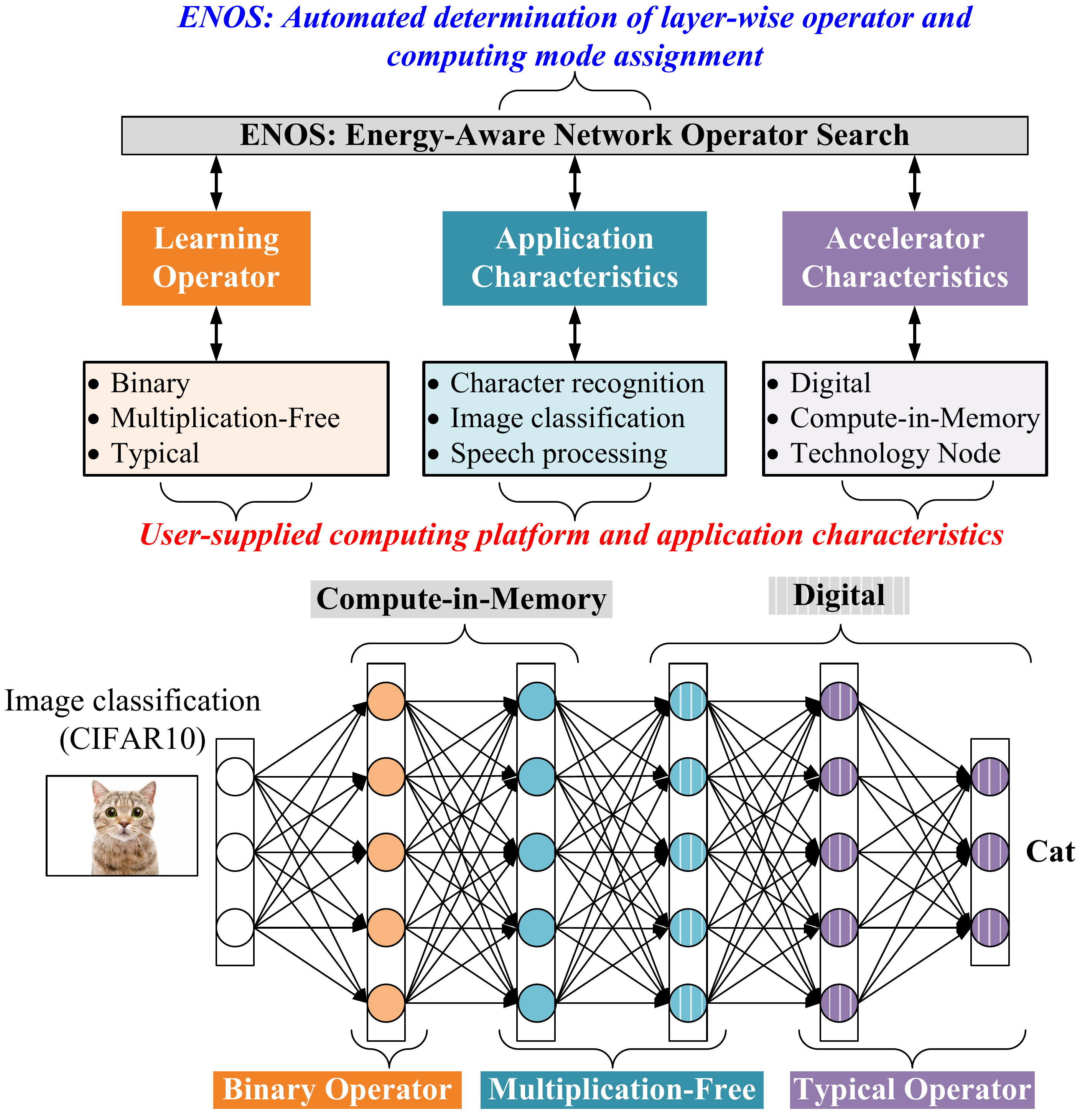}
     \caption{ENOS searches for an optimal layer-wise operator and computing mode assignment to maximize energy efficiency of inference in a DNN while adhering to application and computing platform characteristics such as technology node and compute-in-memory budget.}
     \label{fig1}
\end{figure}
 
Our novel framework on \textit{energy-aware network operator search} (ENOS) synergistically integrates the above two efforts, namely, neural network design automation and exploration of more computationally efficient learning operators. ENOS automatically combines various correlation operators and their computing modes layer-wise for a DNN so that its energy-accuracy trade-off can be optimally balanced. For example, unlike the current approaches, which consider end-to-end DNN processing using a single correlation operator, in Figure \ref{fig1}, ENOS can optimally integrate typical, binary weight, and multiplication-free operators among DNN layers along with their computing modes, such as compute-in-memory or digital. The automated synthesis in ENOS can be made to inherently depend on application and hardware characteristics as well. For example, ENOS can account for a limited compute-in-memory budget in hybrid digital and compute-in-memory accelerators. Similarly, ENOS can recognize task complexity to determine the filter-wise operator sequence in a DNN architecture optimally. Furthermore, inspired by \cite{DART}, ENOS learns in a continuous search space using standard gradient descent-based routines (such as ADAM \cite{kingma2014adam}) so that its training is scalable to complex and large DNNs.

In Sec. II, we discuss the overview of ENOS. In Sec. III, we discuss ENOS on digital platforms while showing results using single-level, bi-level, and sequential optimization strategies. In the section, we also discuss ENOS's variational perspective and demonstrate its performance by sampling filter-wise operator assignments from a learned variational distribution. In Sec. IV, we consider ENOS's application on hybrid accelerators that combine both (low precision) compute-in-memory and (high precision) digital modules. For hybrid accelerators, ENOS is augmented with Lagrangian constraints that prepare the operator and computing mode assignment by considering the limited compute-in-memory budget. Sec. V concludes.   

\begin{figure}[!t]
     \centering
     \includegraphics[width=\linewidth]{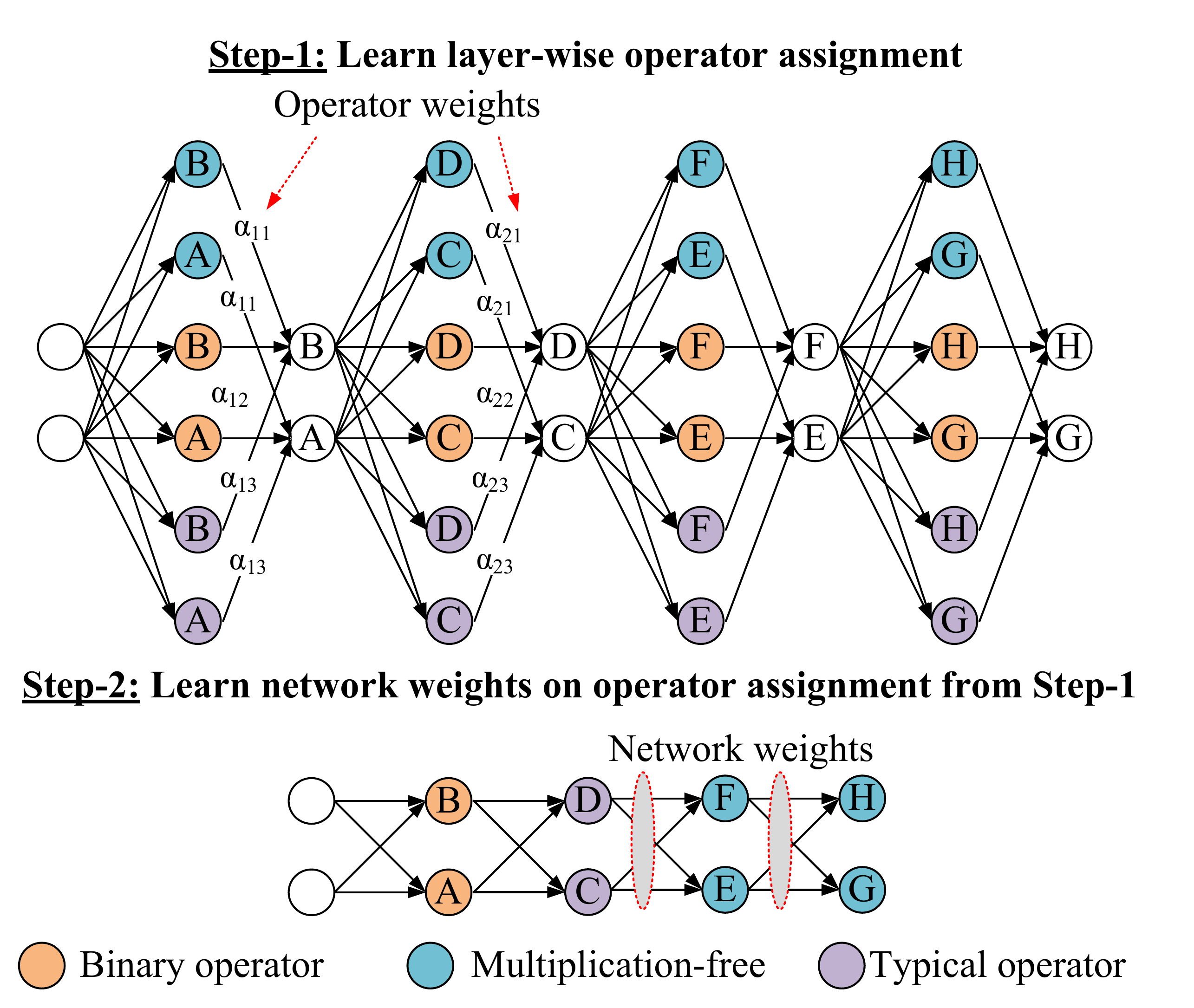}
     \caption{Learning architecture of Single and Bi-Level ENOS: In the first step, at each layer, candidate operators' outputs are combined using linear interpolation with layer-wise operator weight parameters, $\alpha$. The operator weights are also learned through back-propagation. In the second step, only the operators with the highest weight are retained across the network. The corresponding weights for the entire network are relearned. Since the cost function in both steps is continuous, typical learning tools such as ADAM can be utilized.}
     \label{fig2}
\end{figure}

\section{Framework of Energy-Aware Network Operator Search (ENOS)}
%\edit{Useful to see how the paper for \url{https://github.com/zhaoweicai/EdMIPS} is laid out and what theory they highlight.-TT}
Prior works have explored novel deep learning operators such as binarized weights and inputs \cite{courbariaux2016binarized}, binary-connect \cite{binaryconect}, multiplication-free \cite{MFNET}, and deep-shift\cite{deepShift} to improve the computational efficiency of deep learning within limited computing and storage resources. However, computational advantages of simplified operators come at the cost of accuracy degradation. Moreover, the efficiency of novel operators also strongly depends on hardware and task specifications. For example, deep networks based on binarized weights and inputs can be very efficiently implemented using a network of XOR logic gates \cite{simons2019review,rastegari2016xnor}. Despite showing a competitive accuracy on datasets such as MNIST \cite{mnist}, the operator has limited accuracy on more complex datasets such as ImageNet\cite{imagenet_cvpr09} compared to the typical case, i.e., full precision scalar product operator. Therefore, a unique opportunity exists where these operators can be optimally combined, so that their computational efficiency can be harnessed while maintaining high prediction accuracy. Moreover, such optimal layer-wise fusion of diverse network operators should also consider the characteristics of the underlying hardware, energy constraints, and task specifications. Addressing this challenge, we present a systematic framework for such automated operator allocation.

\subsection{Single-level and Bi-level ENOS}

Figure 2 shows the framework of ENOS which is inspired from \cite{DART}. For the ensuing discussion, at each layer, we consider an optimal operator choice among typical, binary connect, and multiplication-free operators. However, our approach is generalizable to any number of and different operator choices. At each layer, input feature map is processed against all three operators across three parallel paths. The output of all three operators are combined using an affine interpolation using the corresponding weight factors ($\alpha$), where the weight factors itself are learning parameters. The overall cost function of DNN is, thus, defined as
\begin{subequations}
\begin{equation}
\mathcal{L}\textsubscript{net} = \underbrace{\mathcal{L}\textsubscript{acc}(\theta,\alpha)}_\text{Accuracy} + \lambda\underbrace{\sum_{i=1}^{N}N_{OP,i}\sum_{j=1}^{M}\text{softmax}(\alpha_{ij})E_{OP,j} }_\text{Regularizer for DNN accelerator energy}
\end{equation}
\begin{equation}
\theta^*,\alpha^* = \text{argmin}(\mathcal{L}\textsubscript{net})
\end{equation}
\end{subequations}
Here, the first term of $\mathcal{L}\textsubscript{net}$, i.e., $\mathcal{L}\textsubscript{acc}(\theta,\alpha)$ minimizes with the higher prediction accuracy while the second term minimizes with the net lower energy of operation. $i$ indexes over network layers, while $j$ indexes over operator choices for each layer. $\theta$ are the network weight parameters. $\alpha$ are the layer-wise operator weight parameters. $\lambda$ is a user-defined hyper-parameter that considers energy-accuracy trade-off. $N_{OP,i}$ is the number of multiply-accumulate (MAC) operations at layer \textit{i}. $E_{OP,j}$ is average energy for an operator \textit{j} for unit operation. 

Without the following non-linear activation (such as ReLU), the operation of input ($\mathbf{x}$) and weight ($\mathbf{w}$) vectors with the three chosen operators, typical, multiplication-free, and binary, are as following, respectively,
\begin{subequations}
\begin{equation}
    f_T(\mathbf{x},\mathbf{w}) = \sum x_i\cdot w_i     %(Typical)
\end{equation}
\begin{equation}
 f_M(\mathbf{x},\mathbf{w}) = \sum \text{sign}(x_i)\cdot \text{abs}(w_i) + \text{sign}(w_i)\cdot \text{abs}(x_i)  %(MF)
\end{equation}
\begin{equation}
 f_B(\mathbf{x},\mathbf{w}) = \sum x_i\cdot binarize(w_i)   %(BNN)
\end{equation}
\end{subequations}
Here, $\cdot$ is an element-wise multiplication operator, $+$ is element-wise addition operator, and $\sum$ is vector sum operator. $\text{sign}()$ operator is $\pm 1$ and $\text{abs}()$ operator produces absolute unsigned value. 

Although the binary operator in Eq. (2c) is non-differentiable, prior works \cite{bengio2013estimating, yang2017bmxnet, peters2018probabilistic} have presented approaches such as straight-through estimators to circumvent the problem. Similarly,  sign() and abs() functions are non-differentiable in Eq. (2b). Prior work \cite{akbacs2015multiplication} approximates the derivative of abs() with a steep hyperbolic tangent and the derivative of sign() with a steep Gaussian function to address this. Therefore, with the added modifications, the net loss function can still be optimized using typical gradient descent approaches such as ADAM \cite{adam}. In the first step of a baseline approach to ENOS, following the above loss formulation, a candidate locally optimal parameter set $\theta^*$ and $\alpha^*$ can be extracted using gradient descent. Then, in the second step, a new design of DNN is considered where only the operator with the highest weight factor ($\alpha_j$) is considered at each layer. The corresponding weights for new DNN are then relearned.

In addition to this single-level baseline optimization approach to simultaneously extract optimal parameter set $\theta^*$ and $\alpha^*$, we also consider a bi-level optimization procedure for step-1 of the learning procedure. Algorithm 1 describes the bi-level optimization procedure, which trains the network hierarchically. On the top level, we update the weight factors $\alpha$ by descending the validation loss $\Delta\mathcal{L}\textsubscript{net}_{Val}$ defined on a dataset separate from training. Keeping the learned $\alpha$, we descend $\Delta\mathcal{L}\textsubscript{net}_{Train}$ to learn the DNN weights $\theta$. Likewise, we iterate through the validation and training data set sequentially to learn the final $\alpha$. 

\begin{algorithm}[]
\SetAlgoLined
\textbf{Goal}: Extract trained $\theta^*$ and $\alpha^*$ to reduce the loss for Step-1.
Loss function: $\mathcal{L}\textsubscript{net} = \mathcal{L}\textsubscript{acc}(\theta,\alpha) + \lambda\sum_{i=1}^{N}N_{OP,i}\sum_{j=1}^{M}\text{softmax}(\alpha_{ij})E_{OP,j}$ 
\\
 \textbf{Given}: Network weights: $\theta$; Operator weights: $\alpha$. \\
 \While {not converged do}{
 1. Update $\alpha$ by descending the loss $\Delta \mathcal{L}\textsubscript{net}_{Val}(\theta^t,\alpha)$\\
 2. Update $\theta$ descending the loss $\Delta \mathcal{L}\textsubscript{net}_{Train}(\theta,\alpha^t)$;
 }
For Step-2, choose the operator with the highest weight factor ($\alpha_j$) at each layer. Relearn the network weights.
 \caption{Bi-Level ENOS}
\end{algorithm}

\begin{figure}[!t]
     \centering
     \includegraphics[width=\linewidth]{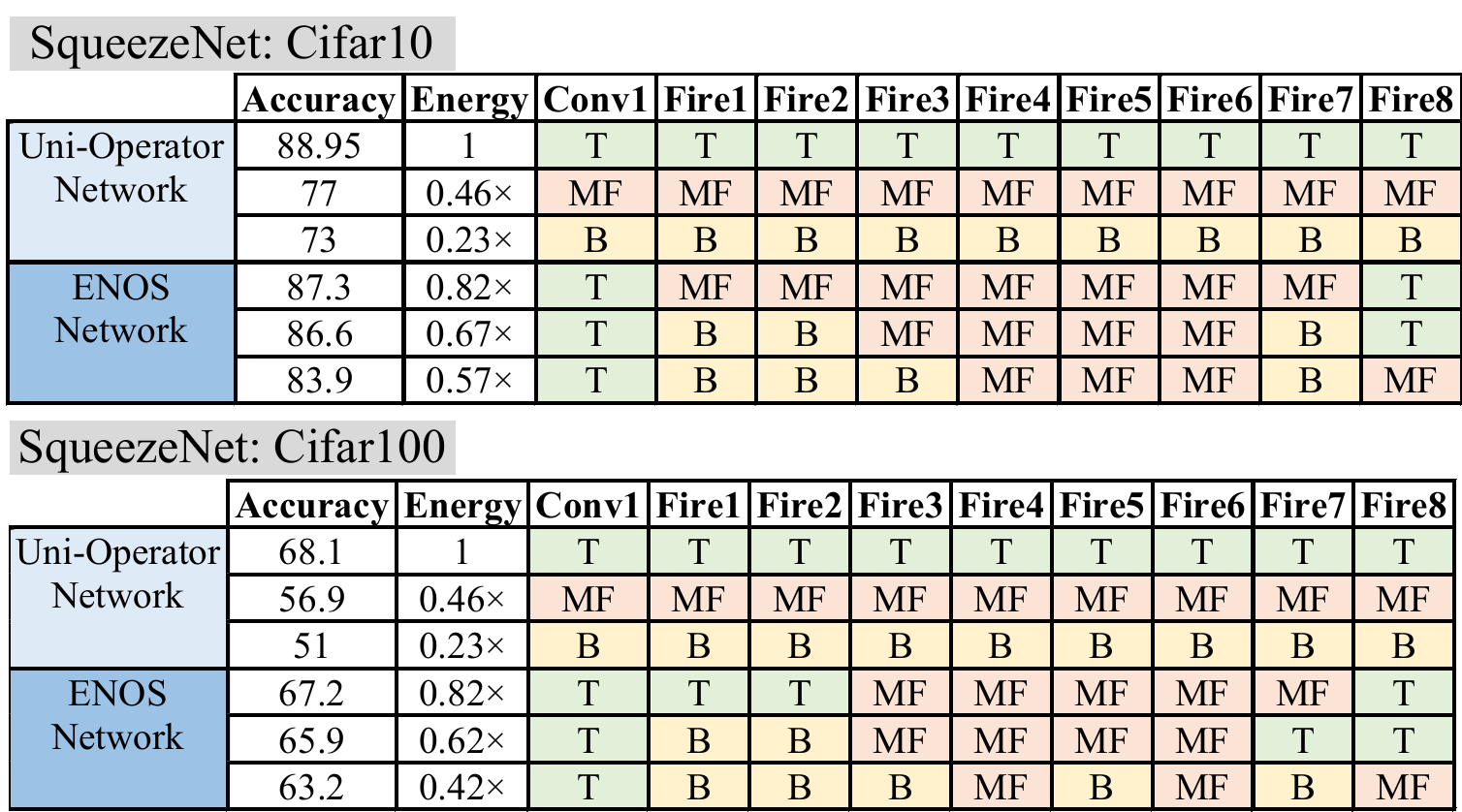}
     \caption{Layer wise operator mapping for SqueezeNet network for Cifar10 and Cifar100 data set. The higher hyper parameter $\lambda$ in equation 1 prioritizes energy efficient operators. The table compares the energy of SqueezeNet with different operator combination.}
     \label{layer}
\end{figure}

Figure \ref{layer} shows an exemplary synthesis of SqueezeNet using ENOS on CIFAR10 and CIFAR100 datasets. More details on ENOS parameters such as extraction of operator energy ($E_{OP,j}$), implementation hardware, and comparison among different optimization objectives are discussed later. Rows in the table show optimal operator assignments under varying $\lambda$. As discussed earlier, increasing $\lambda$ allows trading-off accuracy against DNN accelerator's operating energy. For CIFAR10, compared to typical (T) operator on all layers, optimally integrating different operators on different layers allows $\sim$20\% lower accelerator energy in ENOS at only $\sim$1.5\% accuracy reduction. The lower rows in the figure show even more aggressive energy scaling while suffering moderate accuracy degradation. The application of ENOS on the network also offers many interesting insights. With increasing $\lambda$ to improve energy efficiency, operator assignments begin progressively changing only in the middle convolution layers from typical to multiplication-free and then to binary. Meanwhile, operator assignments for the initial and last convolution layers remain typical. This characterization illustrates that feature extraction in the middle layers in the network is more amenable to simplification and accelerator energy saving without a considerable effect on the overall network's output accuracy.

\subsection{Sequential ENOS}
In the previous approach, concurrent searching of operators for all layers is susceptible to overfitting. To address this concern, we also investigated a sequential mode of operator search. Under the sequential search, operator assignments are selected only for a single layer in one iteration while keeping the other layers' operator choices open. Specifically, a layer ``$i$'' is assigned the operator choice ``$j$'' if it has the highest weight factor $\alpha_{ij}$ among all layers and operator choices. All operator choices are considered in the following training iteration for the remaining unassigned network layers, and their corresponding operator weights are relearned. The iterations continue until all layer-wise operator assignments have been found. Algorithm 2 describes such sequential operator search procedures in ENOS. 

\begin{algorithm}[]
\SetAlgoLined
\textbf{Goal}: Sequentially select the operator for network layers.

Loss function: $\mathcal{L}\textsubscript{net} = \mathcal{L}\textsubscript{acc}(\theta,\alpha) + \lambda\sum_{i=1}^{N}N_{OP,i}\sum_{j=1}^{M}\text{softmax}(\alpha_{ij})E_{OP,j}$ 
\\
 \textbf{Given}: Network weights: $\theta$; Operator weights: $\alpha$. \\
 \For{layer $l$=1 to $L$}{
 \While {not converged}{
Update $\alpha$, $\theta$ by descending the loss on training set $\Delta \mathcal{L}\textsubscript{net}_{Train}(\theta,\alpha)$\\
 }
1. Find the highest network weight factor ($\alpha_{ij}$). \\
2. Assign operator $j$ to layer $i$ and prevent further optimization of this choice. 
 }
Relearn the network weights.
 \caption{Sequential ENOS}
\end{algorithm}

\begin{figure}[!t]
     \centering
     \includegraphics[width=0.9\linewidth]{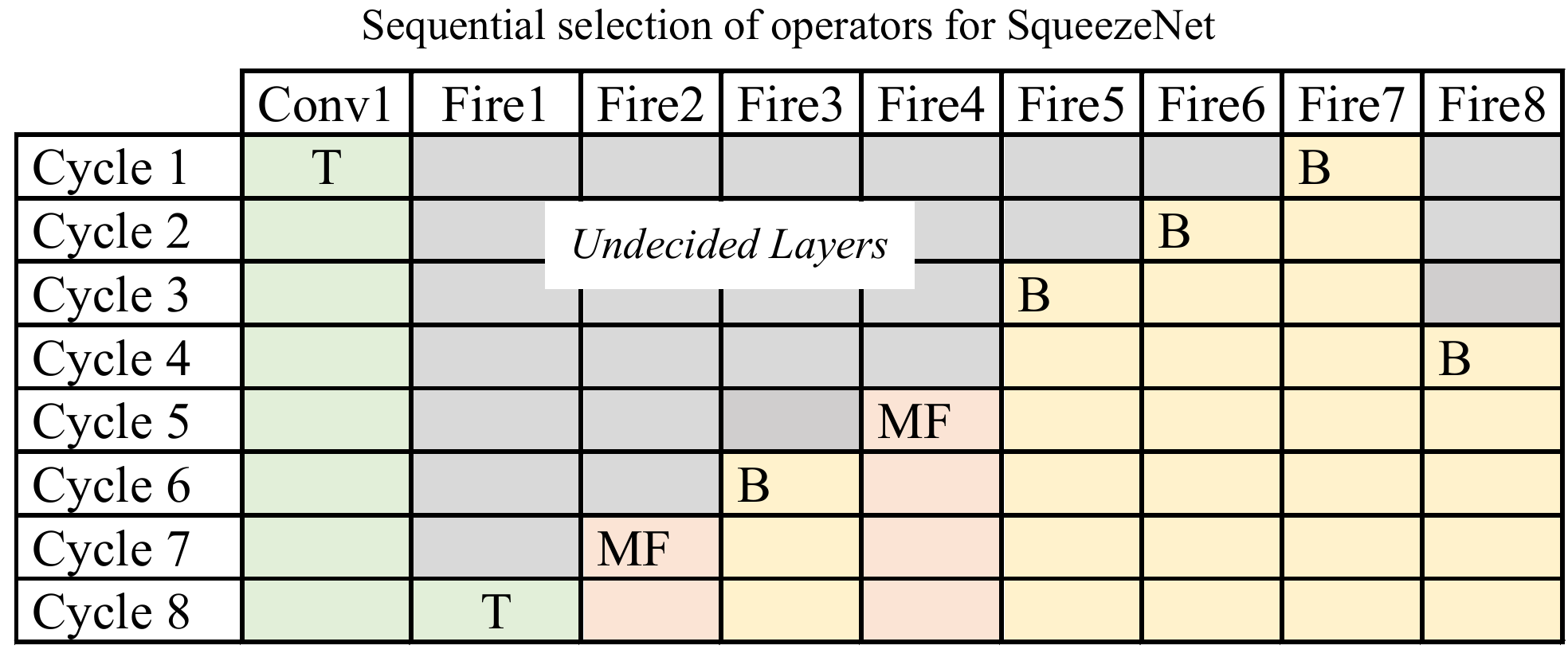}
     \caption{Evolution of operator assignments on the layers of SqueezeNet under sequential optimization in ENOS.}
     \label{seq}
\end{figure}

Figure \ref{seq} shows the evolution of operator assignments on SqeezeNet under sequential ENOS. We kept the operator for the first convolution layer (Conv1) as typical (T) since, under bi-level ENOS, the layer preferred standard operator even under aggressive $\lambda$ scaling. Among the remaining Fire layers, Fire7 is first assigned binary (B) operator since it showed the highest propensity towards B operating under accelerator energy scaling constraints among all the other layers. The operator choices remained open for the other layers. Subsequently, Fire6 is assigned B operator, and Fire1 is assigned T operator at the last cycle. Compared to single-level and bi-level ENOS, the training workload for sequential ENOS is substantially more. Note that the number of training iterations in sequential ENOS scale proportionally to the number of network layers.  

\subsection{Variational ENOS}
The optimal operator assignment problem in ENOS can also be treated under a variational framework. Specifically, softmax($\alpha_{ij}$) can be seen as the parameters of a multinomial distribution that represents the operator choices. Unlike the previous optimization procedures where a layer ``$i$'' is assigned the operator ``$j$'' if $\alpha_{ij}$ is the maximum, the assignments under the variational setting is made by treating softmax($\alpha_{ij}$) as the probability of the optimal operator assignment. Here, a population of $N$ candidate networks can be prepared by sampling the optimal operator choice at each layer ``$i$'' based on the corresponding softmax($\alpha_{ij}$), representing the probability of the operator choice ``$j$'' on the layer. Subsequently, the best performing network from the candidate networks can be selected by testing it on the validation set. Algorithm 3 describes this procedure formally. We note in passing that while one could define a formal variational objective, lower bound it and maximize that with respect to the parameters of the operator choice distribution using the reparametrization trick, we found that our simpler strategy already provides tangible improvements and is closer to the algorithmic templates of single-level, bi-level and sequential strategies.

\begin{algorithm}
\SetAlgoLined
\textbf{Goal}: Learn multinomial distribution parameters representing optimal operator choices and operator assignments. \\
\textbf{Loss}: $\mathcal{L}\textsubscript{net} = \mathcal{L}\textsubscript{acc}(\theta,\alpha) + \lambda\sum_{i=1}^{N}N_{OP,i}\sum_{j=1}^{M}\text{softmax}(\alpha_{ij})E_{OP,j}$ 
\\
 \textbf{Given}: Network weights: $\theta$; Operator weights: $\alpha$. \\
 \textbf{Step-1}: learn $\theta^*$ and $\alpha^*$\\
 \While {not converged}{
 Update $\alpha$, $\theta$ by descending the loss $\Delta \mathcal{L}\textsubscript{net}_{Train}(\theta,\alpha)$
 }
 \textbf{Step-2}: Create a population of candidate networks\\
 \For{samples $n$=1 to N}{
 \For{layer $l$=1 to L}{
    Assign operator based on multinomial distribution parameters softmax($\alpha_{ij}$).
 }
 Relearn network weights} 
 \textbf{Step-3}: Find the optimal operator assignment\\
 \For{samples $n$=1 to N}{
 Characterize the network on validation set.}
 Assign the most accurate network as the optimal choice.
\caption{Variational ENOS}
\end{algorithm}

\begin{figure}[!t]
     \centering
     \includegraphics[width=0.9\linewidth]{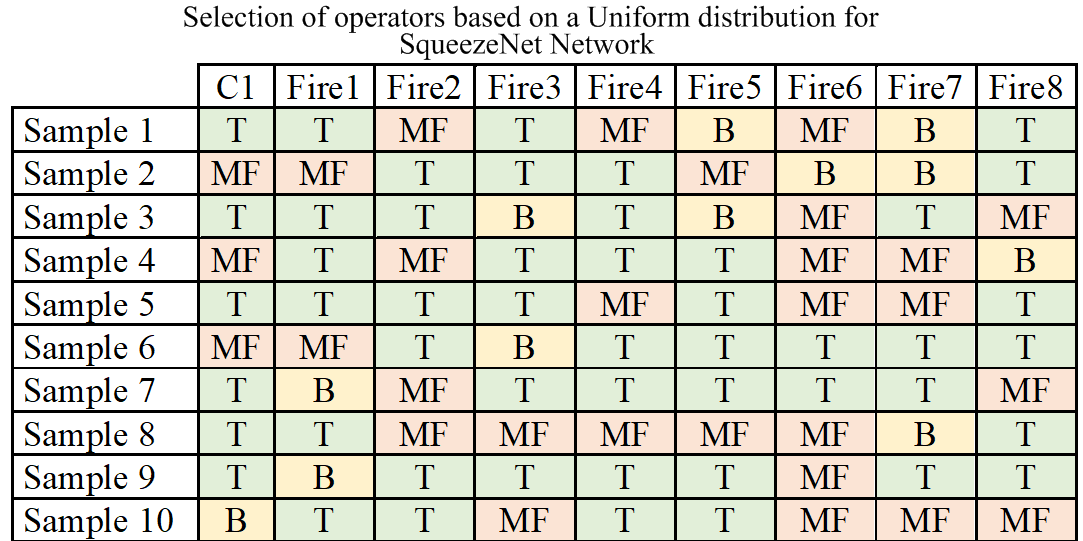}
     \caption{Candidate networks under variational treatment of ENOS. For different layers, the operators are sampled based on the learned multinomial distribution parameters.}
     \label{var}
\end{figure}

Figure \ref{var} shows the operator assignments on various candidate networks under variational ENOS. Interestingly, under the considered $\lambda$, i.e., DNN accelerator's energy constraints, layers Fire4 and Fire6 show less entropy and mostly opt for T and MF operator respectively. Meanwhile, layers such as Fire2 and Fire7 show much higher entropy in their operator assignment. Layer-wise operator assignment among various candidate networks are also jointly correlated as the network searches for the best operator combinations to maximize accuracy under accelerator energy constraints.  

\section{ENOS-based DNN Mapping for Digital Accelerators}

\subsection{Digital MAC unit with reconfigurable operators}
\begin{figure}[t!]
     \centering
     \includegraphics[width=\linewidth]{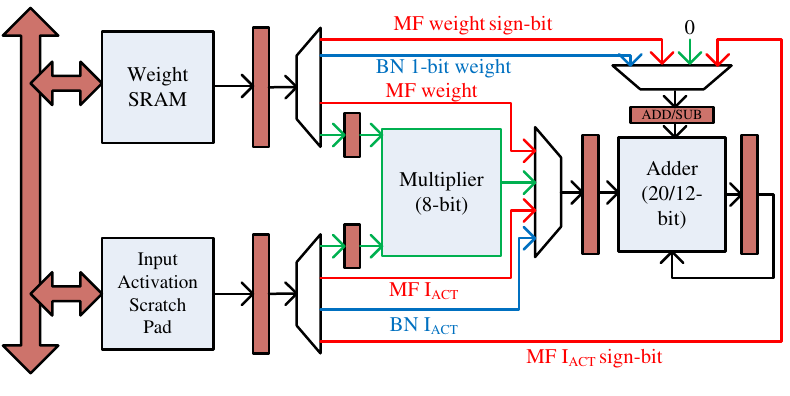}
     \caption{Digital MAC unit with reconfigurable operators. The datapaths for different operators are color coded: datapaths for typical, multiplication-free \& binary operators are highlighted with green, red \& blue, respectively.}
     \label{Reconfigurable_core}
\end{figure}

This section discusses the energy efficiency benefits of ENOS-based DNN mapping considering a digital spatial accelerator. Spatial digital accelerators are prevalent for high-performance DNN \cite{eyeriss,iliev2020low,kim2020energy}. A spatial accelerator combines many parallel multiply-accumulate (MAC) cores using a network-on-chip (NOC). DNN workload is distributed among the parallel cores so that the entire inference flow can be processed with the least latency. Considering ENOS-based DNN mapping on a spatial accelerator, Figure \ref{Reconfigurable_core} shows a MAC unit that can be reconfigured among the three operators that we considered above. A spatial DNN accelerator can combine such reconfigurable MAC cores to flexibly implement ENOS-based layer-wise operator integration under various energy-accuracy constraints.   

In Figure \ref{Reconfigurable_core}(a), datapaths for different operators are associatively color-coded: datapaths for typical, multiplication-free \& binary operators are highlighted with green, red and blue, respectively. Although training for the network weights is performed considering a floating-point precision, an 8-bit fixed precision operation is considered during inference. Similar to prior work \cite{1st,2nd,compac}, 8-bit fixed-precision inference suffers minimal accuracy degradation but can considerably simplify the hardware implementation and improve computations' energy efficiency. Datapath to implement typical operator includes an 8-bit multiplier (to obtain the product of 8-bit input activation and weight) and a 20-bit adder (to accumulate the 16-bit outputs of multiplier continually). Meanwhile, the multiplication-free operator requires just an adder/subtractor block as it essentially implements addition/subtraction of input activation (I$_{ACT}$) with corresponding weight (W), where I$_{ACT}$ and W are both 8-bits. Unlike typical and multiplication-free operators, weights in the binary operator are just one-bit while I$_{ACT}$ is represented using 8-bits. Hence, the binary operator adds/subtracts consecutive input activation depending on corresponding weight bits, therefore, it needs an adder/subtractor block. We use a 12-bit adder with multiplication-free and binary operators since they require a lower dynamic range of I$_{ACT}$ than typical operator. In the considered MAC unit, the adder's dynamic range can be configured to either 20-bit or 12-bit based on a switch. 

The design of the above MAC core was synthesized using Cadence RC Compiler. Predictive technology models (PTM) for 15nm CMOS technology \cite{sinha2012exploring} were used for the synthesis at nominal supply voltage. Timing-closure of the MAC core was obtained considering a 100 MHz clock. Table I shows the energy of a unit operation for all three operators. Typical operator is the most computationally expensive whereas the binary operator requires $\sim$9$\times$ lower compute energy than the typical operator. 

\begin{table}[t!]\vspace{2mm}
\caption{Per operation energy of different operators with 8-bit digital implementation at 1 GHz operating frequency.}
\begin{tabular}{|c|c|c|}
\hline
Operator            & Components                      & Energy/op. (fJ) \\ \hline
Typical             & 8-bit Multiplier + 20-bit Adder & 295.7             \\ \hline
Multiplication-Free & 12-bit Adder                    & 64                \\ \hline
Binary              & 12-bit Adder                    & 32                \\ \hline
\end{tabular}
\label{Energy_numbers}
\end{table}

\begin{figure}[!t]
     \centering
     \includegraphics[width=\linewidth]{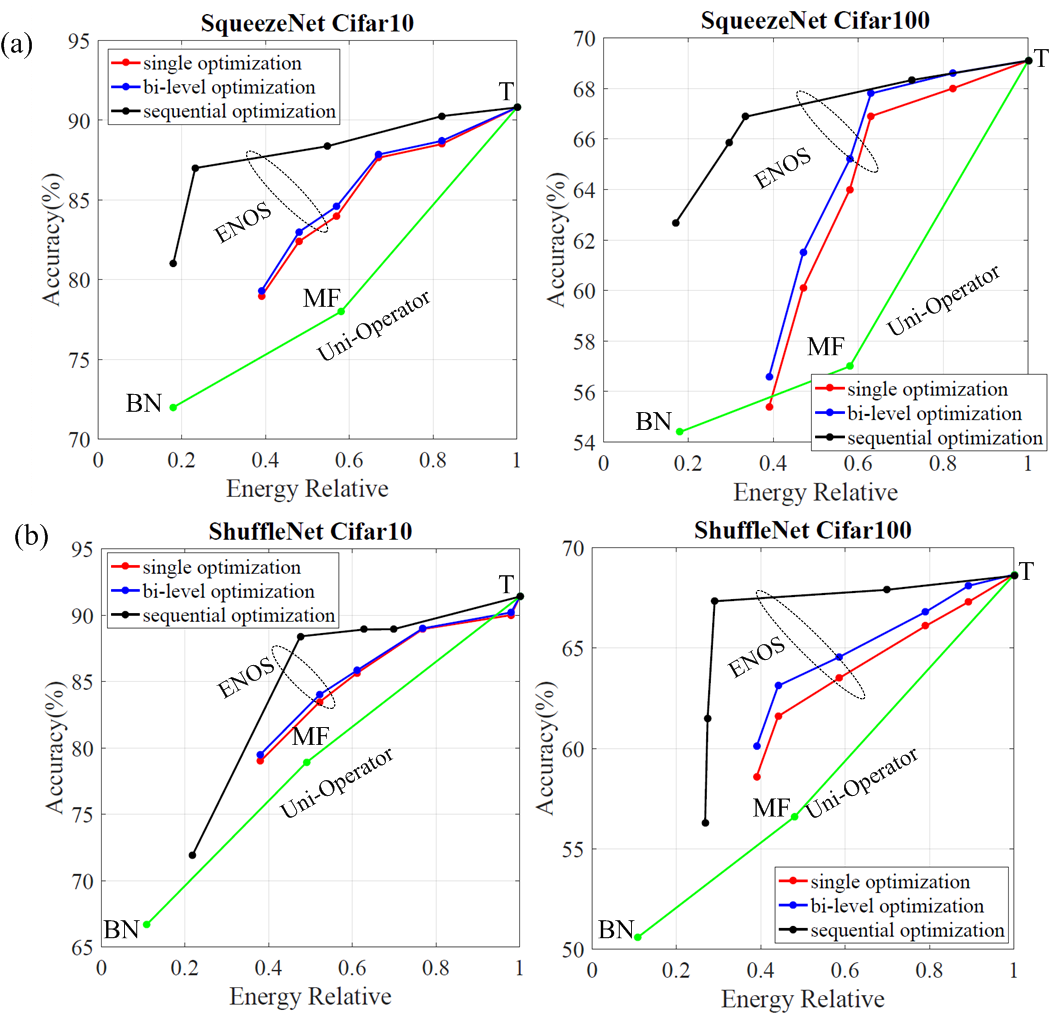}
     \caption{Energy vs. Accuracy on CIFAR10 and CIFAR100 for (a) SqueezeNet and (b) ShuffleNet under single, bi-level, sequential modes of ENOS.}
     \label{energy}
\end{figure}

\subsection{Simulation results and discussions}
We study ENOS on popular networks SqueezeNet\cite{squeezenet} and ShuffleNet\cite{shufflenet}. Note that unlike the predecessor DNNs such as VGG16 \cite{simonyan2014very} and GoogleNet \cite{szegedy2015going}, SqueezeNet and ShuffleNet are already optimized for computational efficiency and edge computing, therefore are harder test-cases for further downscaling of their computing energy. The networks were trained on CIFAR10 and CIFAR100 to search the optimal operator assignments in a layer-wise manner. Figure \ref{energy} compares single-level, bi-level, and sequential operator search modes of ENOS. Different DNN configurations are estimated based on the operator energies in Table I and the number of operations per operator to process the input. For better clarity, the figure shows normalized energies of different configurations under the baseline case where all operations are done using the typical (T) operator.

In Figure \ref{energy}, bi-level ENOS outperforms single-level ENOS under the same energy constraints. For single-level optimization, learning network weights $\theta$ and operator weights $\alpha$ on the same dataset, i.e., training set, makes the training vulnerable to over-fitting. Meanwhile, bi-level optimization learns $\theta$ on the training set and $\alpha$ on the validation set, thereby improves inductive biases by considering more data. Especially, bi-level optimization shows significant improvement in accuracy on a more challenging datasets, i.e., on CIFAR100 compared to CIFAR10. 

Figure \ref{energy} also compares single-level and bi-level ENOS against sequential ENOS. Compared to the concurrent operator assignment modes (single and bi-level), sequential ENOS significantly improves the network synthesis. For example, for SqeezeNet on CIFAR100 [Figure \ref{energy}(a)], sequential ENOS results in a network configuration that has $>$10\% accuracy than single and bi-level ENOS when the energy budget is only 40\% of the baseline all typical operator case. Single and bi-level ENOS modes can only gracefully scale down energy to 60\% of the baseline case; meanwhile, sequential ENOS incurs only $\sim$5\% accuracy loss when the accelerator energy is constrained to be one fifth of the baseline case. Similarly, for ShuffleNet on CIFAR100 [Figure \ref{energy}(b)], sequential ENOS suffers a minimal degradation in energy even when the network energy is scaled down to about 30\% of the baseline case. Here as well, single and bi-level modes of ENOS cannot scale down the network energy to such lower levels. Nonetheless, due to its sequential mode of operator search, sequential ENOS is much more computationally expensive than single and bi-level modes. Significantly, under sequential ENOS, the training iterations grow in sequential ENOS as the number of layers increases.           

Energy-accuracy characteristics of various ENOS modes are also compared to traditional uni-operator cases. DNN accelerator's energy scaling using only binary (B) or multiplication-free (MF) operators in the traditional uni-operator approach considerably degrades the prediction accuracy. Comparatively, with ENOS, optimal integration of various operators allows scaling down accelerator energy to the half without a significant loss in accuracy on all tested cases. If the energy constraints of DNN accelerators are moderate, single and bi-level ENOS have a matching performance to sequential ENOS, albeit, with much less training cost. However, under considerable energy constraints, sequential ENOS surpasses other optimization modes.   

\begin{figure}[!t]
     \centering
     \includegraphics[width=\linewidth]{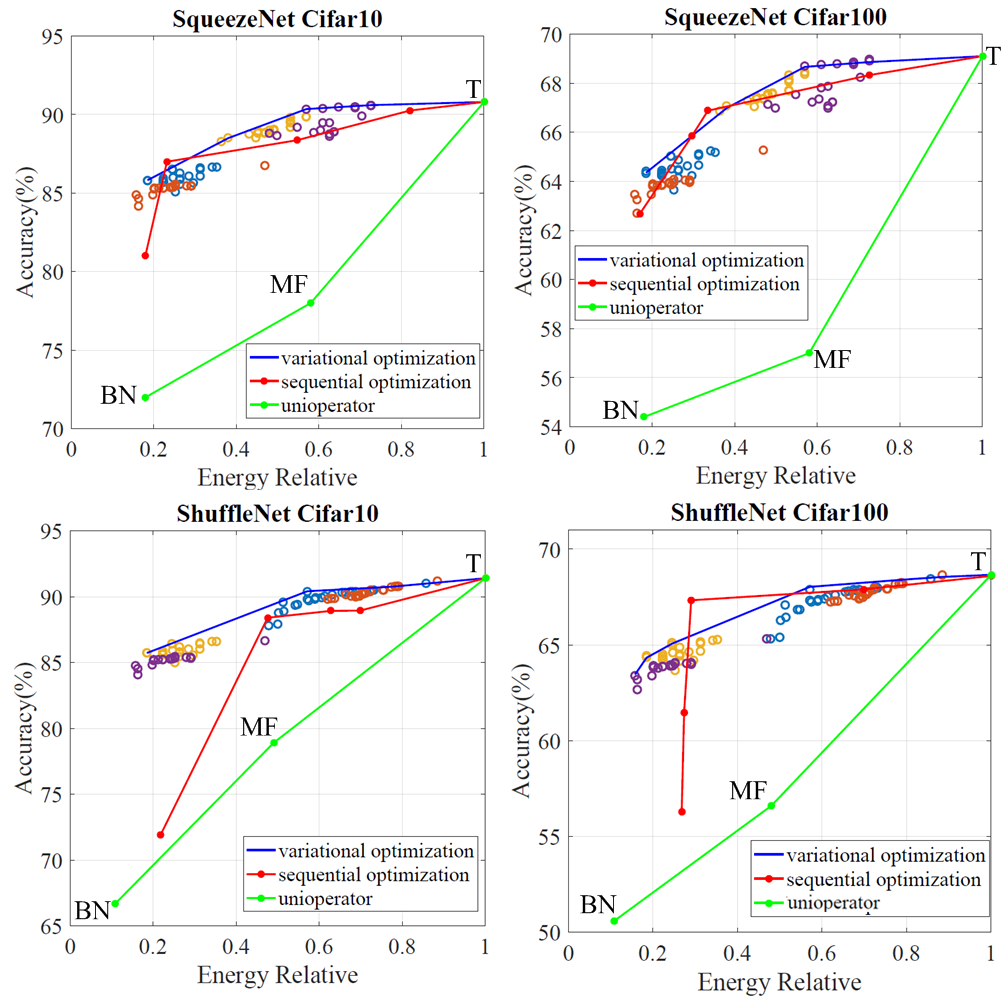}
     \caption{Energy vs. Accuracy on CIFAR10 and CIFAR100 for (a) SqueezeNet and (b) ShuffleNet under variational mode of ENOS and comparison with sequential optimization mode.}
     \label{energy2}
\end{figure}

\begin{figure*}[!t]
     \centering
     \includegraphics[width=\linewidth]{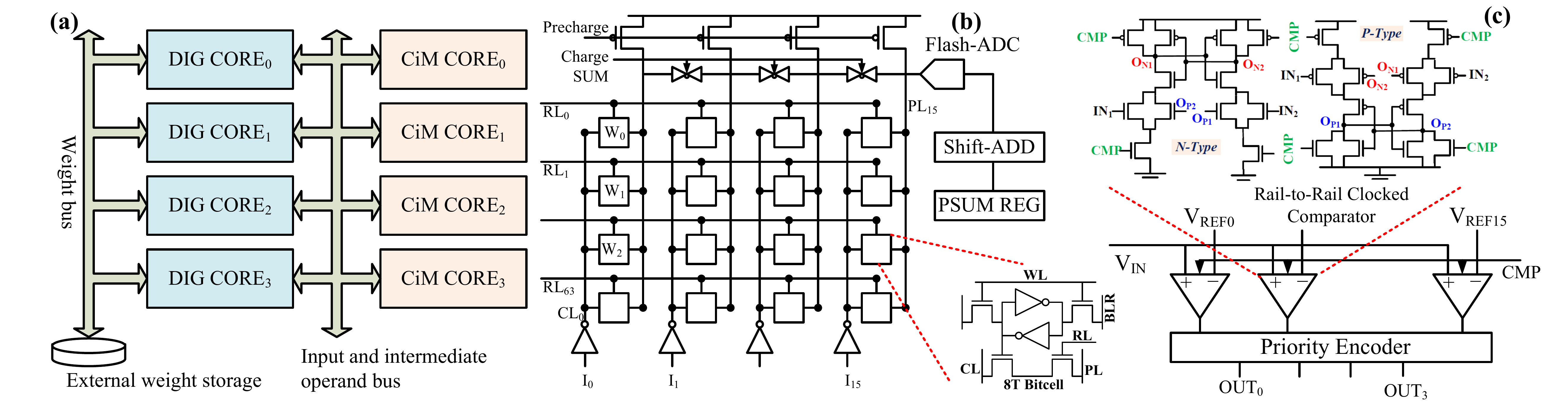}
     \caption{(a) Hybrid DNN accelerator integrating compute-in-memory and digital processing units. (b) Compute-in-memory array with 8-T SRAM cell, Flash-ADC, and digital Shift-ADD unit. (c) Peripherals of compute-in-memory array, rail-to-rail clocked comparator and priority encoder, to digitize output.}
     \label{hybrid}
\end{figure*}

Figure \ref{energy2} shows the characterization of ENOS under variational optimization. In the figure, color-coded clusters of accuracy-energy values are shown corresponding to the candidate networks synthesized by variational ENOS under various $\lambda$ [Algorithm 3]. As we discussed before, the candidate networks are synthesized by treating operator weight softmax($\alpha_{ij}$) as the probability of optimal operator as sampling from a multinomial distribution of operators. Clusters with various $\lambda$ settings correspond to various energy constraints when determining the optimal operator assignments. The blue curve in the figure shows the accuracy-energy trajectory corresponding to the best sample in each cluster, i.e., the one with the highest accuracy. 

Interestingly, in the figure, variational ENOS shows a quite matching performance to sequential ENOS. Under extreme energy constraints, variational ENOS degrades more gracefully and provides better accuracy in the synthesized networks than sequential ENOS. For example, in ShuffleNet, even when the operating energy constraints are $\sim$20\% of the baseline all typical operator case, variational ENOS maintains a graceful degradation. Meanwhile, the accuracy of sequential ENOS drops considerably. Compared to sequential ENOS, variational ENOS is also more computationally efficient. In sequential ENOS, the number of training iterations grows with the number of deep learning layers since each iteration determines the operator for one layer. In variational ENOS, the entire network is trained only once, similar to ENOS's single-level optimization method. Subsequently, many candidate networks are characterized on the validation set, which is a much less workload than training itself. Therefore, even with many candidate samples, variational ENOS maintains high computational efficiency over sequential ENOS.     

\section{ENOS-based DNN Mapping on Hybrid Digital and Compute-in-Memory Accelerators}

\subsection{Hybrid Digital and Compute-in-Memory Accelerator}
To downscale DNN's processing energy even further, mutually-suited hardware modules to computationally efficient learning operators are being investigated. In particular, compute-in-memory (CiM) is becoming a dominant approach in emerging DNN accelerators \cite{verma2019memory}. While a crucial bottleneck in traditional digital DNN accelerators is limited bandwidth between processor and memory, CiM minimizes weight and operand movements by collapsing DNN's parameter storage and computations within the same physical structure. Even more, multiplication-free and binary operators are especially more suited for CiM\cite{MFNET,valavi2018mixed} and allow further reduction in their operating energy. However, since typical CiM units require mixed-signal data conversion circuits (ADC/DAC) integrated within memory modules, CiM's energy and area efficiency is optimal only at a lower precision. Therefore, it is typical to explore CiM-based DNN acceleration under low precision \cite{verma2019memory,biswas2018conv,nasrin2020supported}. 

\begin{table}[]
\centering
\caption{Per operation energy of different operators with 4-bit compute-in-memory (CIM) implementation at 1 GHz operating frequency.}
\label{tab:my-table}
\begin{tabular}{|c|c|c|c|}
\hline
Operator            &  Typical & Multiplication-free & Binary \\ \hline
Energy/op.          &  51.78fJ   & 12.95fJ & 6.47fJ          \\ \hline
\end{tabular}
\end{table}

\begin{figure}[!t]
     \centering
     \includegraphics[width=\linewidth]{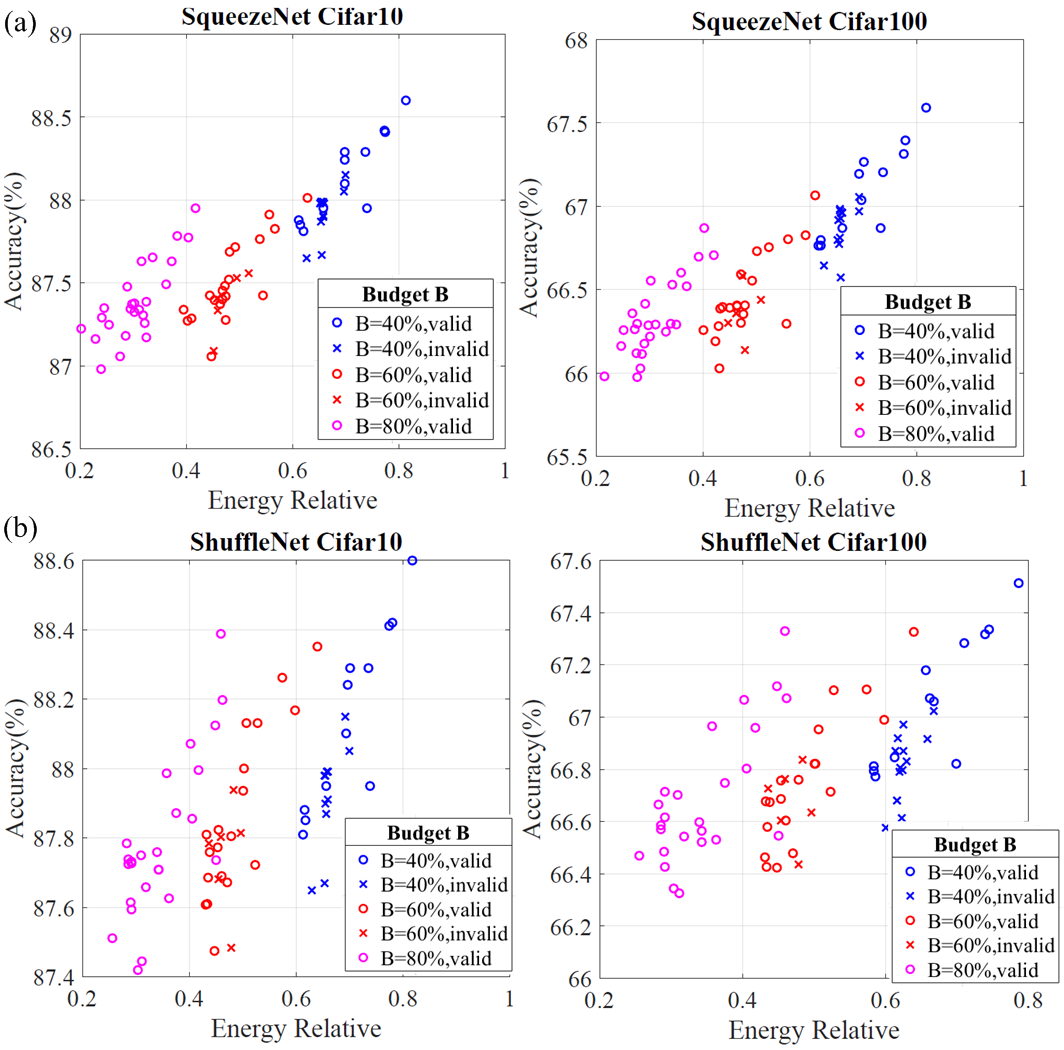}
     \caption{Energy vs. Accuracy of various candidate networks on CIFAR10 and CIFAR100 for (a) SqueezeNet and (b) ShuffleNet under variational mode of ENOS and hybrid digital and compute-in-memory acceleration.}
     \label{inmem}
\end{figure}

Considering the complementary advantages and challenges of digital and CiM for deep learning, in Figure \ref{hybrid}(a), we consider a hybrid DNN accelerator which combines both processing modes and operating precision, higher precision digital units and lower precision mixed-signal compute-in-memory units, and uses ENOS to map DNN layers between them adaptively. In Figure \ref{hybrid}(b), the considered CiM unit is composed of 8-T SRAM cells [shown in the inset of the figure] and uses bitplane-wise processing that avoids DAC as shown in \cite{biswas2018conv}. In the bitplane-wise processing, bitplanes of the same significance input and weight vectors are processed in one cycle. Each SRAM cell has three access ports--RL, PL, and CL--dedicated for CiM operations. Row-lines (RL) connect horizontally in Figure \ref{hybrid}(b) to select the respective row of the array. Product lines (PL) and column lines (CL) connect vertically, PL to the right and CL to the left of a cell. Within memory product is computed on PL while CL applies input patterns.

The CiM module's operation begins with precharging PL and applying input to CL in one half of a clock cycle. Next, RL is activated to compute within cell product bit (zero or one) in one half of a clock cycle. Then, PLs are summed using MUXes in Figure \ref{hybrid}(b). Finally, the sum-line output is digitized using a Flash-ADC. The output of Flash-ADC in consecutive cycles is accumulated using a Shift-ADD unit to determine the net product sum of multi-bit input and weight precision. For Flash-ADC, we use a clocked rail-to-rail comparator as shown in Figure \ref{hybrid}(c). A similar in-memory compute of DNN was discussed in details in \cite{nasrin2021mf}. 

Table II shows the energy of various operators in CiM mode. The energy of various operators are extracted by SPICE simulations using predictive technology models (PTMs) of 15 nm CMOS technology. As we discussed before, due to integrated mixed-signal units, CiM implementations are suitable for low precision operation. Therefore, a 4-bit operation of operators is considered here. Compared to digital processing, CiM-based operations reduce the necessary processing energy by $\sim$5-6$\times$. 

\subsection{ENOS for Hybrid Accelerators}
For such hybrid acceleration, ENOS searches for \textit{mixed-precision}, \textit{mixed-operator}, as well as \textit{mixed-mode} operation of DNN layers to optimize accuracy-energy trade-off. Since, under footprint constraints, a hybrid accelerator may only support limited CiM capacity, the loss function is modified using a Lagrangian to integrate the constraint as  
\begin{equation}
\mathcal{L}\textsubscript{net} = \mathcal{L}\textsubscript{acc}(\theta,\alpha) + \mathcal{L}\textsubscript{en}(\alpha) + \gamma \Big(B - \sum_{i=1}^{N}\sum_{j=1}^{M}\text{softmax}(\alpha_{ij}) \times A_j \times N_{W,i}\Big)^2 
\end{equation}
Here, $\mathcal{L}\textsubscript{acc}$ corresponds to the first term in the right-hand side of (2), and $\mathcal{L}\textsubscript{en}$ corresponds to the energy-dependent second term in (2). $\gamma$ is the Lagrangian parameter that enforces the total weight mapping on CiM units equal to $B$. $i$ is an index over DNN layers, and $j$ is an index over the tuple of operator and processing mode. $\alpha_{i,j}$ is the learned weight of the operator and processing mode tuple. $A_{j}$ is the area cost of per weight for operator $j$. For example, the area-cost per weight in binary operator is 1-bit but in multiplication-free operator is 4-bit in the considered 4-bit implementation in Table II. 

In our previous discussion, variational ENOS was found to be more efficient over other learning modes. Therefore, variational ENOS is also extended for hybrid accelerators. Since Lagrangian-based constrained optimization doesn't guarantee meeting the equality constraints, each candidate network is tested if CiM constraints are satisfied and invalid samples are filtered ouot. Figure 10 shows the exemplary clusters of accuracy-energy for various candidate DNNs, color-coded with the corresponding CiM budget, under variational ENOS for hybrid accelerators. The samples that violate CiM constraints are marked with cross symbol ($\times$). Eligible candidate networks that meet CiM budget are selected and characterized on the validation set by relearning their corresponding weights. The best candidate network with the highest accuracy on validation set is chosen as the final output.  

\begin{figure}[!t]
     \centering
     \includegraphics[width=\linewidth]{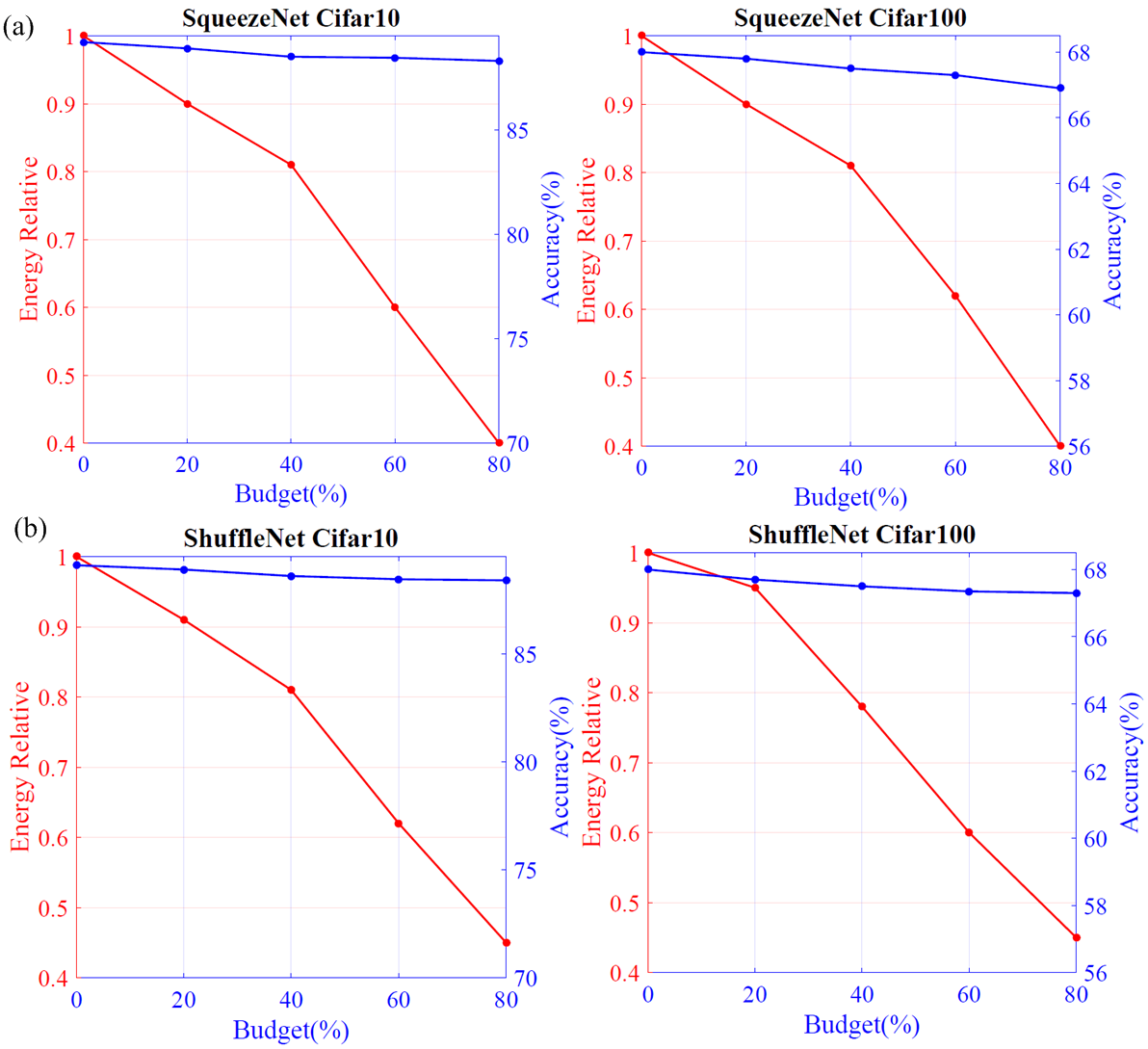}
     \caption{Energy vs. Accuracy on CIFAR10 and CIFAR100 for (a) SqueezeNet and (b) ShuffleNet under varying compute-in-memory budget.}
     \label{inmem1}
\end{figure}

\begin{figure}[!t]
     \centering
     \includegraphics[width=\linewidth]{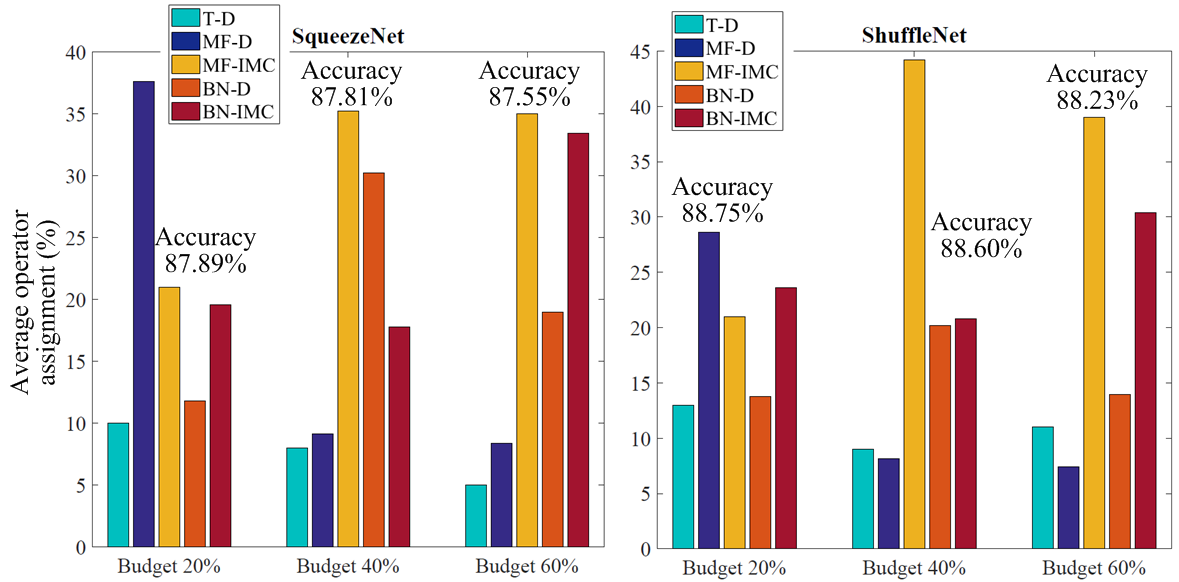}
     \caption{Energy vs. Accuracy on CIFAR10 and CIFAR100 for (a) SqueezeNet and (b) ShuffleNet under varying compute-in-memory budget.}
     \label{inmem2}
\end{figure}

\subsection{Simulation Results and Discussions}
Figure \ref{inmem1} shows the energy vs. accuracy trends of variational ENOS under hybrid acceleration. We consider many hybrid accelerator configurations of varying CiM budget. CiM budget of an accelerator is expressed in terms of the percentage of weights that can be operated with CiM mode [using modules in Figure 9] while the remaining weights need to be operated digitally [using modules in Figure 6]. An accelerator with higher CiM capability can perform most of the operation in highly energy efficient CiM mode, but also requires more area to implement since CiM units are not amenable to multiplexing unlike digital units. In Figure 12, with increasing the CiM budget, ENOS adaptively allocates more CiM units to exploit their higher energy efficiency while staying within the allocated CiM budget. Interestingly, compared to the digital accelerator's variational ENOS in Figure 8, variational ENOS on the hybrid accelerator can rather exploit the flexibility of selecting operating mode, rather than operator choices alone, to allow a significant reduction in operating energy with only a minimal accuracy loss. 

\section{Conclusions}
This paper presents a novel class of efficient energy-aware operator search algorithms (ENOS) for a wide variety of deep neural networks. ENOS can match the state-of-the-art performance metrics for neural networks on image classification with remarkable energy efficiency improvement, and can be applied as-is to other domains/applications such as Transformer based language models. Additionally, we present a reconfigurable MAC core that implements the chosen optimal operators by carefully considering the energy-accuracy trade-offs. In the future, the layer-level operator search approach presented here can be expanded to neuron-level operator search. The ENOS approach can also be widened to incorporate other similar goals in multiple resource constrained settings beyond energy-efficiency.

\bibliographystyle{IEEEtran}
\bibliography{main.bib}
%\addbibresource{ref.bib}
\end{document}